\documentclass[fleqn,10pt]{wlscirep}

\title{Apparent diffusion in nucleus pulposus is associated with pain and mobility improvements after spinal mobilization for acute low back pain} 

\author[1]{Paul Thiry}
\author[1,+]{Fran$\c c$ois Reumont}
\author[2]{Jean-Michel Brism\'{e}e}
\author[3,4,+,*]{Fr\'{e}d\'{e}ric Dierick}
\affil[1]{OMT Skills, Private physical therapy practice, La Louvi\`{e}re, 7100, Belgium}
\affil[2]{Center for Rehabilitation Research and Department of Rehabilitation Sciences, Texas Tech University Health Sciences Center, Lubbock, Texas, USA}
\affil[3]{Forme \& Fonctionnement Humain Research Unit, Physical Therapy Department, Haute Ecole Louvain en Hainaut, Montignies-sur-Sambre, 6061, Belgium}
\affil[4]{Universit\'{e} catholique de Louvain, Faculty of Motor Sciences, Louvain-la-Neuve, 1348, Belgium}

\affil[*]{frederic.dierick@gmail.com}

\affil[+]{these authors contributed equally to this work}


\begin{abstract}
Pain perception, trunk mobility in flexion, extension, and lateral flexion, and apparent diffusion coefficient ($ADC$) within nucleus pulposus of all lumbar discs were collected before and after posterior-to-anterior mobilization in 16 adults with acute low back pain. $ADC$ was computed from diffusion maps and 3 specific portions of the nucleus pulposus were investigated: anterior ($ADC_{ant}$), middle ($ADC_{mid}$), and posterior ($ADC_{post}$), and their mean as $ADC_{all}$, a summary measure of $ADC$ within nucleus pulposus. 
Pain ratings were significantly reduced after mobilization, and mobility of the trunk was significantly increased. Concomitantly, a significant increase in $ADC_{all}$ values was observed. The greatest $ADC_{all}$ changes were observed at the L$_{3}$-L$_{4}$ and L$_{4}$-L$_{5}$ levels and were mainly explained by changes in $ADC_{ant}$ and $ADC_{post}$.
The simultaneous reduction in pain and increase of water diffusion within nucleus pulposus has has been previously observed in subjects with chronic conditions and exists in the acute phase of the disease. Since the largest changes in $ADC$ were observed at the periphery of the nucleus pulposus, and taken together with pain decrease, our results suggest that increased peripheral random motion of water molecules is implicated in the modulation of the intervertebral disc nociceptive response.
 \end{abstract}

\begin{document}
\flushbottom
\maketitle
%
%
\thispagestyle{empty}

\noindent 

\section*{Introduction}
Among all musculoskeletal pain conditions, the prevalence and burden from low back pain (LBP) [ICD-10-CM, code M54.5] is very high throughout the world: out of the 291 conditions studied in the Global Burden of Disease 2010 study, LBP ranked highest in terms of disability and sixth in terms of overall burden \cite{Hoy:2014}. Spinal mobilization is a very common approach for LBP, and when a spinal mobilization is correctly performed by a trained orthopaedic manual physical therapist (OMPT), the intervention has low risk of injury and may result in immediate detectable improvements in pain and larger articular amplitudes. However, despite the widespread use of lumbar joint mobilization, the physiological responses of lumbar anatomical structures are still largely unknown. Recent advances in magnetic resonance imaging (MRI) of the musculoskeletal have nevertheless allow to observe the movement of water within and between tissues \textit{in vivo}, and is called diffusion-weighted (DW) MRI. This emerging imaging technology is particularly sensitive to small changes in fluid flow and has a great potential for studying the influence of physical therapy interventions such as manual therapy, exercise, and physical agents on musculoskeletal structures \cite{Beattie:2011}. Based on the comparison between DW images and non-DW images using the same MRI sequence, it is possible to reconstruct the mapping of the diffusion and to calculate an apparent diffusion coefficient ($ADC$) within intervertebral disc (IVD) \cite{Kerttula:2000, Kerttula:2001, Antoniou:2004, Newitt:2005}. Interestingly, DW MRI of the IVD has been successfully used for some years by Beattie and his colleagues \cite{Beattie:2008, Beattie:2009, Beattie:2010, Beattie:2014} and allowed to link the decreasing pain reported by subjects with chronic LBP following single session of lumbar posterior-to-anterior (PA) pressures from L5 to L1 levels associated to McKenzie prone press-ups \cite{McKenzie:1981}, to the increase in $ADC$ values in the lumbar IVD \cite{Beattie:2010} or high-velocity, short-amplitude thrust at L5-S1 level \cite{Beattie:2014}. From a physiological point of view, diffusion of water within IVD has been suggested as one mechanism of analgesia following manual mobilization/ manipulation \cite{Beattie:2011}, but the complete mechanism is still unknown. 

Despite the exciting and innovative natures of the studies that explored simultaneously $ADC$ in IVD and pain changes after spinal mobilization/ manipulation in LBP patients, different methodological choices may have influenced the results and make it difficult to generalize to a real clinical setting. First, it consisted of prescribed mobilization/ manipulation in a population, including LBP patients with heterogeneous chronicity and intensity of symptoms \cite{Beattie:2010}. Second, only young patients were included, with a mean around 35 years in a study \cite{Beattie:2010} and even younger (around 25 years) in the other \cite{Beattie:2014}. Since Wu \textit{et al.} \cite{Wu:2013} showed that significant higher $ADC$ values in young asymptomatic subjects (age $<$45 years) are observed at each IVD lumbar level compared to elderly (age $>$45 years), studies including older subjects with chronic LBP are now necessary to generalize the results previously observed. Third, $ADC$ values were only computed in the IVD central portion, corresponding to the center of the nucleus pulposus (NP). It is therefore also necessary to compute values in adjacent regions of the central portion of the NP to be able to better understand the global physiological response of the nuclear part.

Today, a more pragmatic trial investigating the effect of spinal mobilizations on $ADC$ of lumbar IVD, pain perception and trunk mobility changes is needed. Indeed, altered general or segmental kinematic behavior of the trunk, whether restricted, excessive, or linked to poor motor control, is associated with LBP \cite{Abbott:2006,Kulig:2007} and their identification frequently guides the conservative therapeutic approach \cite{Hicks:2005,Kulig:2007}. Therefore, we conducted a single arm, nonrandomized quasi-pragmatic pilot trial with the objective to better understand the short-term effect of a unique PA mobilization technique on $ADC$ of lumbar IVD, pain perception and trunk mobility changes in subjects suffering from idiopathic acute LBP. Contrary to previous studies using DW MRI to assess the physiological response of IVD from a single region of interest (ROI), $ADC$ maps were computed in 9 ROIs in the NP and correlations between $ADC$, pain perception and trunk mobility changes were also explored.

\section*{Methods}
\subsection*{Subjects}
A priori estimation of the sample size was carried out by using G*Power software (Version 3.1.9.2), with an $\alpha$ level (I) equal to 0.05 and $\beta$ level (II) equal to 0.20, with a statistical power of 0.80. The estimation was made on the basis of the average results obtained by Beattie \textit{et al.} \cite{Beattie:2014} which have shown a significant increase of the $ADC$ at the L$_{1}$-L$_{2}$ IVD (1.70$\pm$0.25 $\times$ 10$^{-3}$$mm^{2}~s^{-1}$ \textit{versus} 1.80$\pm$0.24 $\times$ 10$^{-3}$$mm^{2}~s^{-1}$) after lumbar PA mobilization in young subjects with LBP with a low pain intensity. An effect size $dz$ of 0.41 was calculated for unilateral t test for paired samples and a correlation between the groups of 0.5. The estimate of the total size of the sample data is 39 that means 20 subjects suffering from acute LBP, as each subject provides two data.

This study was conducted on a sample of 16 adult patients (11 women and 5 men) suffering from acute idiopathic LBP diagnosed by a physician, that were consecutively recruited from a private physical therapy practice (OMT Skills, La Louvi\`{e}re, Belgium); age: 46$\pm$16 years (range: 26-85), height: 165.8$\pm$9 $cm$, weight: 73.4$\pm$17 $kg$, and body mass index: 26.6$\pm$4 $kg~m^{-2}$. The inclusion and exclusion criteria of the subjects were similar to previous studies \cite{Wilder:2011, Cramer:2013}. 
                                                                                                                                                                                                                                                                                                                                                                                                                                                                                                                                                        Inclusion criteria were: be aged between 20 to 85 years, suffering from acute LBP ($<$ 6 weeks of pain), having 1 month without pain between the current and previous episodes of LBP, subject must have had more days without pain than days with pain in the past year. Exclusion criteria include: aversion to spinal manipulation, chronic LBP, radiating pain below the knees, spine fracture or surgery, osteoporosis, pregnancy, implanted devices that could interact with the magnetic field of MRI, claustrophobia, obesity, alcohol or drug abuse, mental illness or lack of cognitive ability. 

The study protocol and the informed consent documents have been approved by the medical ethics committee of the Universit\'{e} catholique de Louvain (2014/07AOU/419) -- Belgian registration nr = B403201421675; reference number on BioMed Central : ISRCTN16069685 DOI 10.1186/ISRCTN16069685. All research was performed in accordance with relevant guidelines/regulations, and informed consent was obtained from all participants

\subsection*{General procedure}
Before participation in the study, all procedures were explained to all subjects, and they signed an informed consent.

One of the investigators (R.F.) invited the subjects to complete a VAS for pain, a DN4 questionnaire, and a shortened version of McGill Pain Questionnaire validated in French (Questionnaire Douleur Saint-Antoine, QDSA) \cite{Bijur:2001, Bourreau:1984, Bouhassira:2005}. QDSA has 58 word descriptors categorized into 16 subgroups, including 9 sensory groups and 7 affective groups. The subjects pick the word descriptors and score them from 0 (not at all) to 4 (extremely). A sensory (QDSA-S), affective (QDSA-A), and total score (QDSA-T) of QDSA was computed as the sum of A to I (/36), J to P (/28), and A to P items (/64), respectively. A second investigator, blinded in relation to the first one's evaluations, invited the subjects to evaluate their pain using an OAS and performed various trunk mobility tests in standing posture: flexion [$TF$], extension [$TE$] and left and right lateral flexion [$TLF_{l}$ and $TLF_{r}$. A neuro-dynamic test in sitting posture, called slump test \cite{Maitland:1985} was also conducted.

A first MRI scan of the lumbar region of the subject was then carried out. After this scan, a spinal Maitland's PA mobilization \cite{Engeveld:2013} was performed by another investigator (T.P.). The mobilization was realized in a consultation room, very close to the scanner, and equipped with a classic medical examination table. A mechanical floor weighing scale (Seca 762, Hamburg, Germany) was placed under the feet of OMPT to note the weight exerted and the change in weight exerted during PA mobilization. At this point in time, neither of the two investigators were informed of the results of the initial imaging. To complete the data collection, a second MRI scan, identical to the first, was carried out on the subject, within an hour after the spinal mobilization. After the second scan, pain ratings and trunk mobility tests were again performed by the two investigators. 

Total time of the procedure was around 90 $min$, including 2 $\times$ 12 $min$ for MRI, and 45 $min$ for physical examination (pain ratings and trunk mobility tests) and questionnaires (subjects were in sitting posture during around 35 $min$).  

\subsection*{Physical examination and PA mobilization}
The physical examination was done by the principal investigator (T.P.), a certified OMPT, with more than 30 years of experience. It consisted of a complete orthopaedic manual therapy physical examination, inspired by Maitland's physical examination \cite{Engeveld:2013}, and aimed to collect information, first subjective (interrogation) and then objective (physical assets), to confirm the origin of the lumbar pain symptoms of the subject. It also allowed the OMPT to reassess the subject after the spinal mobilization. During trunk mobility tests ($TF$, $TE$, $TLF_{l}$, and $TLF_{r}$), a centimetric measure of major fingertip-to-floor distance was made before and after mobilization.

For PA mobilizations, the OMPT chose: the location of force application on the spinous process(es), the components of the movements and the grades (rhythm and amplitude) varying with his feelings and the evolution of the patient's pain \cite{Engeveld:2013, Olsen:2009}, and duration of mobilizations, as during treatment at own office. Total duration of the mobilizations was timed, and primary (more than half the total mobilization time) and secondary (less than half the time) locations of the applied forces on spinous processes were gathered.

\subsection*{MRI acquisition}
Two lumbar MRI scans were realized for each patient, one before and one after spinal mobilization. All sessions were conducted at the same time of the day (6:00--8:00 PM) to control the diurnal variations of the fluid content in IVDs.

The procedure used for image acquisition is similar to the one described by Beattie \textit{et al.} \cite{Beattie:2008}. All images were obtained using a 1.5 Tesla MRI scanner (MAGNETOM Symphony, Siemens AG, Munich, Germany) at the nuclear magnetic resonance department of Grand H\^{o}pital de Charleroi (Site of ``Notre-Dame'', Charleroi, Belgium). Multi-element spine coils were used for the T2-weighted and DW images. An abdominal coil was also used for the DW images. Subjects entered the scanner head first, with the hips and knees flexed to approximately 30 degrees. Spin echo techniques were used to obtain T2-weighted sagittal and axial views using the parameters described in Table \ref{table:parameters}. DW image parameters are also summarized in Table \ref{table:parameters}. For each slice, DW imaging was obtained by applying diffusion gradients in 3 orthogonal directions and the mean $ADC$ was constructed on the basis of averages of signal intensity from 3 directional DW images \cite{Beattie:2008}. The diffusion-weighting $b$-factor was 400 $s~mm^{-2}$, regarded as the best combination of diffusion weighting and signal intensity \cite{Kealey:2005,Beattie:2008,Beattie:2009,Beattie:2014}.

\begin{table}[!h]
\centering
\caption{T2- and DW parameters used for MRI. FoV: field of view; TE: echo time; TR: repetition time.}
\label{table:parameters}
\begin{tabular}{@{}ll@{}}
\toprule
\multicolumn{2}{c}{T2-weighted images}                                                                                                                                                                                                                                                                                                                                                                                                                                                                                                                                                                                       \\ \midrule
\begin{tabular}[c]{@{}l@{}}Slice group : 1\\ Slices : 13\\ Dist. Factor : 10\%\\ Position : R6.3 A23.2 F21.5\\ Orientation : S \textgreater T2.1\\ Phase enc. Dir. : H\textgreater\textgreater F\\ Phase oversampling : 70\%\\ Flip angle : 150 deg\\ Fat suppr. : none\\ water suppr. : none\\ Antennes : SP3 / SP4 / SP5\end{tabular} & \begin{tabular}[c]{@{}l@{}}FoV read : 300 mm\\ FoV phase : 100.0\%\\ Slice thickness : 4.0 mm\\ Base resolution : 384\\ Phase resolution : 75\%\\ TR: 3500 ms\\ TE: 93 ms\\ Averages : 2\\ Concatenations : 1\\ Filter:\\ Distortion corr. (2D)\\ Coil elements : SP3-5\end{tabular} \\
\midrule
\multicolumn{2}{c}{Diffusion-weighted images}                                                                                                                                                                                                                                                                                                                                                                                                                                                                                                                                                                                \\ \midrule
\begin{tabular}[c]{@{}l@{}}Slice group : 1\\ Slices : 16\\ Dist. Factor : 10\%\\ Position : R9.3 P11.6 F61\\ Orientation : S \textgreater T3.6\\ Phase enc. Dir. : A\textgreater\textgreater P\\ Phase oversampling : 34\%\\ Fat suppr. : SPAIR\\ Antennes : SP2 / SP3 / SP4 / SP5\end{tabular}                                         & \begin{tabular}[c]{@{}l@{}}FoV read : 400 mm\\ FoV phase : 100.0\%\\ Slice thickness : 4.0 mm\\ Base resolution : 192\\ Phase resolution : 80\%\\ TR : 3500 ms\\ TE : 88 ms\\ Averages : 4\\ Filter : Distortion corr. (2D)\\ Coil elements : SP3-6\end{tabular}                     \\
\bottomrule                                                                                                                                                                                                                                                                                                                                                                                                                                                                                                                                             
\end{tabular}
\end{table}

\begin{figure}[!h]
\centering
\includegraphics[scale=0.5]{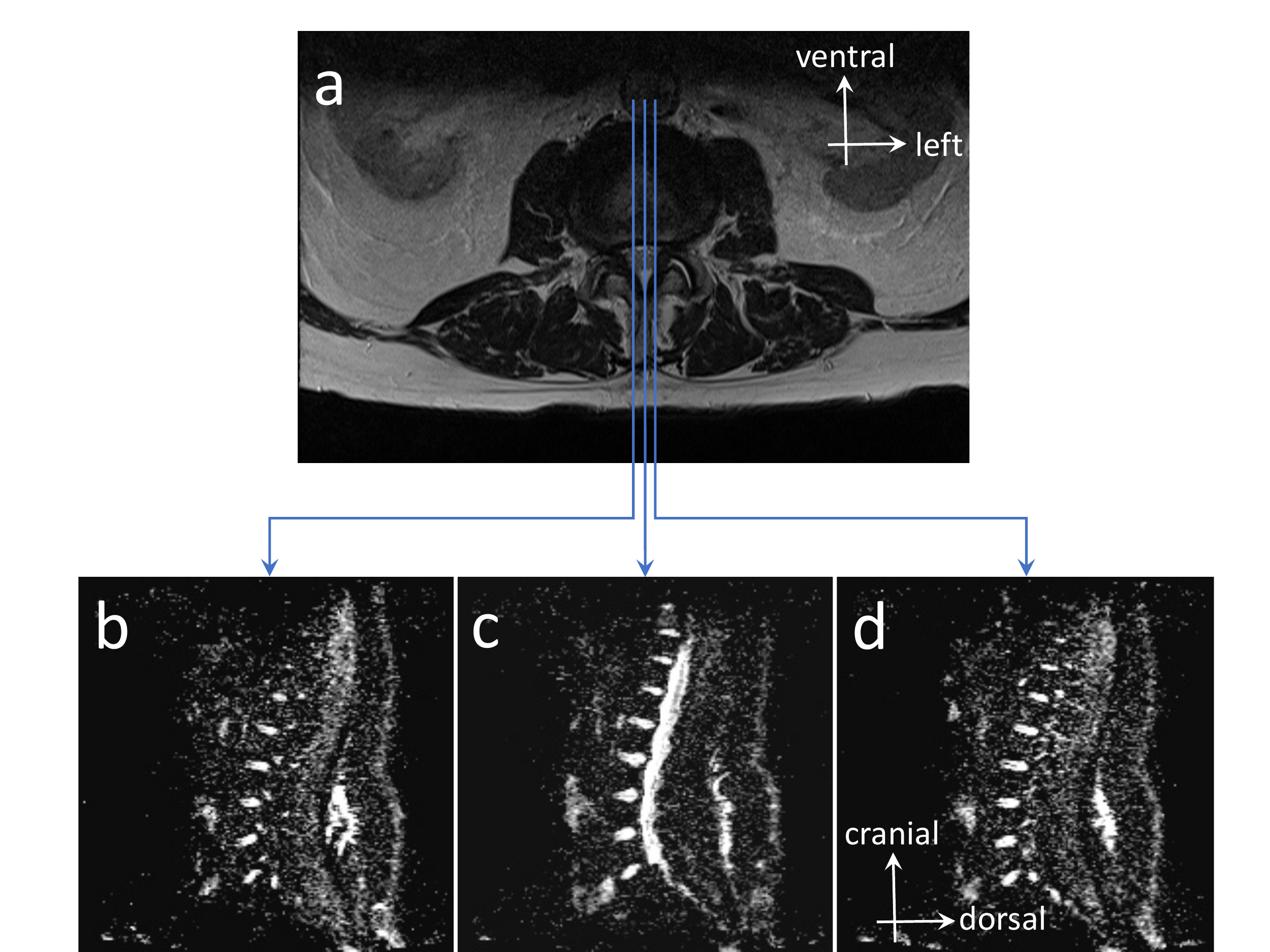}
\caption{T2-weighted MRI cross section at L$_{4}$-L$_{5}$ IVD and $ADC$ mappings in sagittal medial and parasagittal planes. 
Position of the 3 section planes are shown on T2 image (a) and their resultant $ADC$ mappings in parasagittal right (b), sagittal medial (c), and parasagittal left (d) planes.}
\label{fig:method}
\end{figure}

A 3-level modified version \cite{Beattie:2008} of the grading system initially developed by Pfirrmann \textit{et al.} \cite{Pfirrmann:2001} was used to identify the presence and extend of IVD degeneration. Intensity (brightness) and homogeneity of the T2 signal in the nuclear region of midsagittal images was estimated for all IVDs. Hyperintense, homogenous, bright-white NP, with a clear distinction between the AF and NP was graded as 1 (normal); inhomogenous, gray NP, that can be distinguished from the AF as 2 (intermediate); and inhomogenous, gray or black NP, that can not be distinguished from the AF as 3 (hypointense). Each of the T2-weighted images of all subjects were evaluated independently by one of the investigators (R.F.) and a radiologist, with more than 30 years of experience in the field of musculoskeletal system, to classify the IVDs and consensus between the 2 examiners was used to address any disagreements in classification \cite{Beattie:2010}.

\subsection*{Image analysis}
Diffusion sequences were acquired to quantify the micro-movements of water molecules within the IVD of the lumbar spine. $ADC$ was computed and provides the image of the mobility of water molecules. Maps of the mean $ADC$ were calculated on-line by the MRI scanner with the standard software. After the images were obtained, the files were saved and transferred to a remote workstation for analysis.

The interpretation of the images and the calculation of the $ADC$ were achieved by the radiologist and one investigator (R.F.). $ADC$ measurements were conducted for each IVD in the sagittal medial (Figure \ref{fig:method}c), and right and left parasagittal planes (Figures \ref{fig:method}b and \ref{fig:method}d, respectively). The adequate position of the 3 section planes used for $ADC$ measurements were verified by linking them to a T2-weighted cross section passing through the IVD (Figure \ref{fig:method}a).

The adequate position of half-height of each IVD of the lumbar spine on the $ADC$ map was determined using a T2-weighted cross section passing through the IVD. 

$ADC$ were computed from 9 specific ROIs (Figures \ref{fig:balls and methods}a and b) of 0.2 $cm^2$ surface that were selected respectively in the anterior, middle and posterior portions of IVD along the sagittal medial (ROIs \#2, \#5, and \#8) and parasagittal left (ROIs \#1, \#4, and \#7) and right planes (ROIs \#3, \#6, and \#9). Mean of anterior ROIs \#1 to \#3 ($ADC_{ant}$), middle ROIs \#4 to \#6 ($ADC_{mid}$), posterior ROIs \#7 to \#9 ($ADC_{post}$) were computed. Mean of $ADC_{ant}$, $ADC_{mid}$, $ADC_{post}$ was computed as $ADC_{all}$.

\subsection*{Statistical analyses}
All statistical procedures were performed with SigmaPlot software (Version 11.0, Systat Software, San Jose, CA). A one-way RM ANOVA was used to compare the VAS results between before and after the mobilization. A two-way RM ANOVA was used to compare the centimetric data results for bending tests of flexion, extension and left and right lateral flexions and $ADC$ results in IVDs between before and after the mobilization.

All data are presented as means and SD and were checked for normality (Shapiro-Wilk) and equal variance tests. A two-way (level $\times$ treatment) RM ANOVA with a \textit{post hoc} Holm-Sidak method for pairwise multiple comparisons was performed and used to examine the effect of the mobilization by PA pressures. The effect size ($\eta^2$) was calculated as the sums of the squares for the effect of interest (level, treatment and level $\times$ treatment) divided by the total sums of the squares. The significance level $\alpha$ was set at 0.05 for all analyses and \textit{post hoc} statistical power was calculated (SigmaPlot, Version 11.0, Systat Software, San Jose, CA).

To determine whether $ADC_{all}$ correlates with clinical data of pain (VAS) and trunk mobility ($TF$, $TE$, $TLF_{l}$ and $TLF_{r}$), a PCA was performed with R software (\texttt{FactoMineR} and \texttt{factoextra} packages).

Intra-rater reliability of $ADC$ measures realized between 2 sessions by R.F. investigator was determined at one year interval for the paired measures at randomly selected IVD levels of 3 randomly selected subjects. The values obtained were exactly the same, showing perfect intra-rater reliability and therefore not requiring the calculation of an intraclass correlation coefficient ($ICC$).

Test-restest (relative) reliability of $ADC$ measures between 2 MRI scans for one LBP subject (male, 33 years, 183 $cm$, 93 $kg$, pain duration: one week) was estimated using an $ICC$ calculated using R software (\texttt{irr} package), based on a single rater/measurement, absolute-agreement, two-way random effects model (ICC(2,1), see Shrout and Fleiss \cite{Shrout:1979}). The subject was sitting on an chair during 35 $min$ between the 2 measures, and do not receive the lumbar mobilization intervention. Good to excellent relative reliability results were observed with $ICC$ ranging from 0.86 to 0.98.

Within-subject variability, or absolute reliability, attributable to repeated measures between 2 MRI scans, was assessed by the standard error of measurement percent change ($SEM_{\%}$) calculated as (SEM/Mean) $\times$ 100, where SEM is the standard error of measurement and Mean is the mean of all observations from the 2 scans. SEM was calculated as SD $\times$  $\sqrt{1-ICC}$ ~\cite{Beattie:2008}, where SD is the standard deviation of the pooled measures of the 2 scans. $SEM_{\%}$ results ranged from 2.1 to 4.7.

\section*{Results}
\subsection*{Classification of T2-weighted signal of nuclear region}
Percentage of subjects for the 3 grades on the modified Pfirrmann grading system were: 0\% for grade 1, 87.5\% for grade 2, and 12.5\% for grade 3 at L$_{1}$-L$_{2}$; 12.5\%, 81.3\%, and 6.2\% at L$_{2}$-L$_{3}$; 18.8\%, 75\%, and 6.2\% at L$_{3}$-L$_{4}$; 12.5\%, 37.5\%, and 50\% at L$_{4}$-L$_{5}$; 6.2\%, 43.8\%, and 50\%, respectively, at L$_{5}$-S$_{1}$.

\subsection*{Clinical data}
Mean$\pm$SD total duration of PA mobilizations was 639$\pm$102 $s$. Primary locations of PA mobilizations were applied at L$_{1}$ ($n$=1), L$_{3}$ ($n$=3), L$_{4}$ ($n$=7), and L$_{5}$ ($n$=5) levels, and secondary locations were only applied on 3 subjects at T$_{11}$ ($n$=1), L$_{1}$ ($n$=1), and L$_{5}$ ($n$=1) levels. All subjects had a DN4 score $<$4, indicating the absence of neuropathic pain. Median (Q1--Q3) QDSA-T was 22 (18.5--26.5), QDSA-S was 13.5 (9.75--16.25), and QDSA-A was 10 (5.75--11.5).

VAS and OAS pain ratings were significantly reduced after mobilization with a very large effect size (Table \ref{table:one and two way}). A mean$\pm$SD reduction on VAS of 3.4$\pm$1.7 on 10 (62$\pm$25\%) was observed. Mobility of the trunk, assessed by $TF$, $TE$, $TLF_{l}$, and $TLF_{r}$, was significantly increased with medium to large effect sizes (Table \ref{table:one and two way}). A mean reduction of major fingertip-to-floor distance of 6 $cm$ was observed for $TF$, 5 $cm$ for $TE$, 4 $cm$ for $TLF_{l}$, and 5 $cm$ for $TLF_{r}$.

\begin{table}[!h]
\centering
\caption{One-way RM ANOVA results for pain and trunk mobility. $VAS$: visual analogue scale; $OAS$: oral analogue scale; $TF$: trunk flexion; $TE$: trunk extension; $TLF_{l}$: lateral flexion left; $TLF_{r}$: lateral flexion right; significant values are in bold. Two-way RM ANOVA results for $ADC$, stratified according to IVD level and location. Results are expressed in units of 10$^{-6}$ $mm^{2}~s^{-1}$. CI: confidence interval; $ADC_{all}$ mean of $ADC_{ant}$, $ADC_{mid}$, and $ADC_{post}$; $ADC_{ant}$: mean of anterior ROIs; $ADC_{mid}$: mean of middle ROIs; $ADC_{post}$: mean of posterior ROIs.}
\label{table:one and two way}
\begin{tabular}{@{}lccccccc@{}}
\toprule
                                    & Before               & After                & F                 & P-value                              & Power              & Effect size  ($\eta$\textsuperscript{2})           \\ \midrule
                                   & Mean$\pm$SD (95\% CI) & Mean$\pm$SD (95\% CI)   &    &    &     &    \\
Pain (on 10)               &            &               &                   &                                      &                    &                  \\
$VAS$                                 & 5.4$\pm$1.9              & 2.1$\pm$1.5              & 61.9              & \textbf{\textless0.001}              & 1.000              & 0.510                   \\
$OAS$                                 & 5.5$\pm$1.6              & 2.3$\pm$1.7              & 61.8              & \textbf{\textless0.001}              & 1.000              & 0.523                   \\

Mobility (cm)              &                      &                      &                   &                                      &                    &                         \\
$TF$                                  & 28$\pm$15                & 19$\pm$13                & 12.9              & \textbf{0.003}                       & 0.911              & 0.092                   \\
$TE$                                  & 62$\pm$5                 & 57$\pm$6                 & 13.2              & \textbf{0.002}                       & 0.919              & 0.199                   \\
$TLF_{l}$                                 & 50$\pm$6                 & 46$\pm$6                 & 20.5              & \textbf{\textless0.001}              & 0.991              & 0.157                   \\
$TLF_{r}$                                 & 49$\pm$8                 & 44$\pm$5                 & 14.3              & \textbf{0.002}                       & 0.939              & 0.130 \\                  
&  &  &  &  &  &  \\
$ADC_{all}$              &   &   &   &   &   &                                              \\
Treatment       &  &           & 98.9 & \textbf{\textless0.001} & 1.000 & 0.026       \\
Level           &  &           & 6.4  & \textbf{\textless0.001} & 0.971 & 0.208       \\
Level $\times$ Treatment & &           & 5.7  & \textbf{\textless0.001} & 0.944 & 0.006       \\
L$_{1}$-L$_{2}$               & 1437$\pm$233 (1188--1685)                                & 1536$\pm$231 (1290--1781)     &  &   &   &                                 \\
L$_{2}$-L$_{3}$               & 1477$\pm$196 (1268--1686)                                & 1559$\pm$180 (1367--1751)      &  &  &  &                           \\
L$_{3}$-L$_{4}$               & 1333$\pm$315 (997--1668)                                 & 1493$\pm$297 (1177--1810)      &  &  &  &                           \\
L$_{4}$-L$_{5}$              & 1073$\pm$346 (705--1442)       			& 1223$\pm$333 (869--1577)     &  &  &  &        \\
L$_{5}$-S$_{1}$               & 1210$\pm$356 (830--1589)                                 & 1236$\pm$338 (876--1597)         &  &  &  &                         \\

$ADC_{ant}$              &  &  &  &                                   & \textbf{}                                    \\
Treatment      & &           & 83.8 & \textbf{\textless0.001} & 1.000 & 0.041       \\
Level           & &           & 3.9  & \textbf{0.007}          & 0.755 & 0.143       \\
Level $\times$ Treatment & &           & 4.2  & \textbf{0.005}          & 0.796 & 0.008       \\
L$_{1}$-L$_{2}$               & 1277$\pm$240 (1022--1533)                                 & 1377$\pm$248 (1112--1641)         &  &  &  &                         \\
L$_{2}$-L$_{3}$               & 1320$\pm$202 (1105--1535)                                 & 1445$\pm$190 (1243--1647)         &  &  &  &                        \\
L$_{3}$-L$_{4}$              & 1161$\pm$298 (844--1478)                                 & 1367$\pm$265 (1084--1649)         &  &  &  &                         \\
L$_{4}$-L$_{5}$              & 991$\pm$322 (649--1334)                                  & 1130$\pm$298 (812--1447)          &  &  &  &                        \\
L$_{5}$-S$_{1}$              & 1174$\pm$313 (841--1508)                                 & 1210$\pm$334 (854--1566)         &  &  &  &                        \\

$ADC_{mid}$              & &  &  &                                    & \textbf{}                                    \\
Treatment       & &           & 21.2 & \textbf{\textless0.001} & 0.992 & 0.014       \\
Level          & &           & 7.3  & \textbf{\textless0.001} & 0.988 & 0.226       \\
Level $\times$ Treatment & &          & 4.8  & \textbf{0.002}          & 0.874 & 0.465       \\
L$_{1}$-L$_{2}$               & 1541$\pm$252 (1273--1809)                                 & 1612$\pm$241 (1355--1869)        &  &  &  &                          \\
L$_{2}$-L$_{3}$               & 1599$\pm$173 (1415--1783)                                & 1644$\pm$175 (1458--1831)          &  &  &  &                       \\
L$_{3}$-L$_{4}$               & 1420$\pm$313  (1087--1755)                               & 1567$\pm$274 (1275--1858)          &  &  &  &                       \\
L$_{4}$-L$_{5}$              & 1138$\pm$358 (757--1519)                                & 1290$\pm$358 (909--1671)             &  &  &  &                    \\
L$_{5}$-S$_{1}$               & 1293$\pm$406 (860-1725)                                & 1277$\pm$377 (876--1679)             &  &  &  &                     \\

$ADC_{post}$            &   &  &  &                                    &                                              \\
Treatment       & &           & 69.4 & \textbf{\textless0.001} & 1.000 & 0.022       \\
Level          & &           & 7.1  & \textbf{\textless0.001} & 0.984 & 0.219       \\
Level $\times$ Treatment & &           & 1.9  & 0.121                   & 0.264 & 0.006       \\ 
L$_{1}$-L$_{2}$               & 1492$\pm$273 (1202--1783)                                 & 1618$\pm$243 (1359--1878)           &  &  &  &                       \\
L$_{2}$-L$_{3}$               & 1512$\pm$268 (1227--1797)                                & 1588$\pm$229 (1344--1832)             &  &  &  &                     \\
L$_{3}$-L$_{4}$               & 1416$\pm$368 (1024--1808)                                & 1547$\pm$365 (1158--1936)              &  &  &  &                   \\
L$_{4}$-L$_{5}$               & 1090$\pm$417 (646--1534)                                 & 1250$\pm$381 (844--1656)              &  &  &  &                   \\
L$_{5}$-S$_{1}$               & 1162$\pm$375 (763--1562)                                 & 1221$\pm$340 (859--1583)         &  &  &  &    \\ \bottomrule
\end{tabular}
\end{table}

\subsection*{Diffusion of water within discs}
\begin{figure}[!h]
\centering
\includegraphics[scale=0.5]{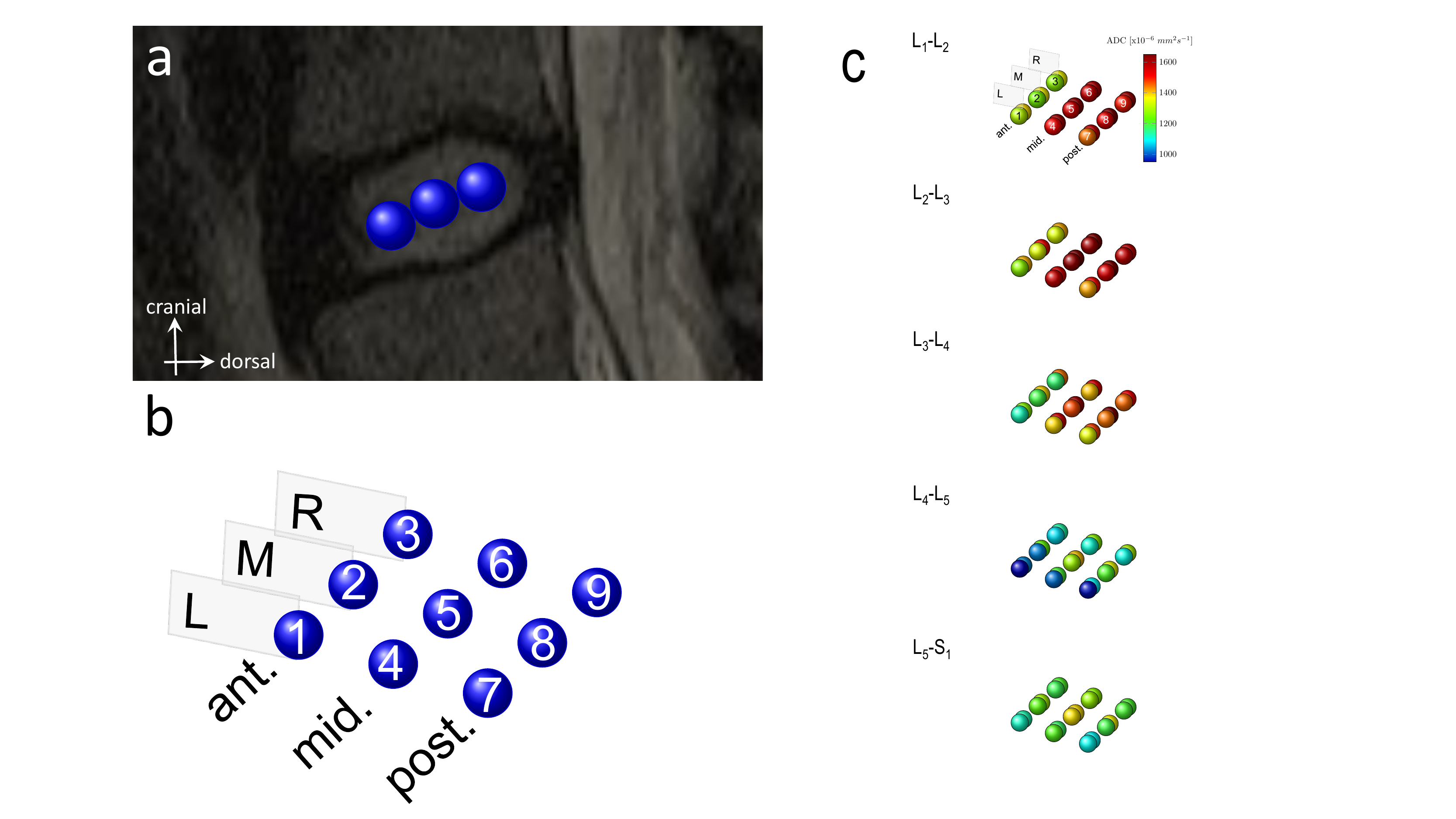}
\caption{(a and b) Nine ROIs studied in each IVD. (c) Mean $ADC$ values before and after intervention, for the 9 ROIs (\#1 to \#9) at the 5 anatomical levels (L$_{1}$-L$_{2}$ to L$_{5}$-S$_{1}$). Anterior (ant.), middle (mid.) and posterior (post.) portions of IVDs along the sagittal medial (M, ROIs \#2, \#5, and \#8), parasagittal left (L, ROIs \#1, \#4, and \#7) and right planes (R, ROIs \#3, \#6, and \#9). Values before the intervention are represented by the circles in the foreground and the ones after the intervention in the background.}
\label{fig:balls and methods}
\end{figure}

Mean $ADC$ values before and after intervention, for the 9 ROIs at the 5 anatomical levels for anterior, middle, and posterior portions of IVDs along the sagittal medial, and parasagittal left and right planes are presented in Figure \ref{fig:balls and methods}.

A significant mean increase in $ADC_{all}$ values was observed after mobilization, with difference of means between 82.1 (change of 5.9\%) and 160.7 $\times$ 10$^{-6}$ $mm^{2}~s^{-1}$ (13.2\%) (Tables \ref{table:one and two way} and \ref{table:posthoc}). Similar significant results were observed in the anterior ($ADC_{ant}$ between 99.2 (8.8\%) and 205.5 $\times$ 10$^{-6}$ $mm^{2}~s^{-1}$ (20\%)), middle ($ADC_{mid}$ between 71.1 (5\%) and 151.8 $\times$ 10$^{-6}$ $mm^{2}~s^{-1}$ (16\%)), and posterior portions of the IVD ($ADC_{post}$ between 76.1 (6.0\%) and 159.8 $\times$ 10$^{-6}$ $mm^{2}~s^{-1}$ (20.1\%)). Significant differences in $ADC_{all}$, $ADC_{ant}$, $ADC_{mid}$, and $ADC_{post}$ were observed at all anatomical levels, except L$_{5}$-S$_{1}$ (Table \ref{table:posthoc}). In addition, no significant difference was observed in $ADC_{mid}$ at L$_{2}$-L$_{3}$ (Table \ref{table:posthoc}). The greatest $ADC_{all}$ changes were observed at the L$_{3}$-L$_{4}$ and L$_{4}$-L$_{5}$ levels and were mainly explained by changes in $ADC_{ant}$ and $ADC_{post}$ (Table \ref{table:posthoc}).

\begin{table}[!h]
\centering
\caption{Post hoc results of two-way RM ANOVA for $ADC$, stratified according to IVD level and location. Difference of means (in units of 10$^{-6}$ $mm^{2}~s^{-1}$) and mean change (\%) in $ADC$ after mobilization. CI: confidence interval; $ADC_{all}$ mean of $ADC_{ant}$, $ADC_{mid}$, and $ADC_{post}$; $ADC_{ant}$: mean of anterior ROIs; $ADC_{mid}$: mean of middle ROIs; $ADC_{post}$: mean of posterior ROIs; significant values are in bold.}
\label{table:posthoc}
\begin{tabular}{@{}lcccc@{}}
\toprule
                            & Difference  &  Change (95\% CI)                     & t                             & P-value                                     \\ \midrule
$ADC_{all}$                      &         &                                &                               &                                             \\
L$_{1}$-L$_{2}$                      & 89.9   & 7.2 (5.3--9.1)                                & 4.3                           & \textbf{\textless0.001}                     \\
L$_{2}$-L$_{3}$                     & 82.1       & 5.9 (3.2--8.6)                            & 3.6                           & \textbf{\textless0.001}                     \\
L$_{3}$-L$_{4}$                      & 160.7         & 13.2 (8.7--17.7)                           & 7.0                           & \textbf{\textless0.001}                     \\
L$_{4}$-L$_{5}$                       & 149.9            & 16.0 (9.6--22.3)                       & 6.6                           & \textbf{\textless0.001}                     \\
L$_{5}$-S$_{1}$                       & 26.5             & 4.1 (-2.4--10.6)                      & 1.2                           & 0.250                                       \\
 &  &  &  &  \\
$ADC_{ant}$                      &         &                                & \textbf{}                     &                                             \\
L$_{1}$-L$_{2}$                       & 99.2          & 8.8 (3.3--14.2)                          & 3.3                           & \textbf{0.001}                              \\
L$_{2}$-L$_{3}$                       & 124.9         & 10.0 (5.9--14.1)                         & 4.2                           & \textbf{\textless0.001}                     \\
L$_{3}$-L$_{4}$                       & 205.5    & 20.0 (13.3--26.8)                              & 6.8                           & \textbf{\textless0.001}                     \\
L$_{4}$-L$_{5}$                       & 138.2       & 16.6 (7.5--25.6)                           & 4.6                           & \textbf{\textless0.001}                     \\
L$_{5}$-S$_{1}$                       & 35.7         & 3.9 (-3.0--10.8)                           & 1.2                           & 0.238                                       \\
&   &  &  &  \\
$ADC_{mid}$                     &           &                              & \textbf{}                     &                                             \\
L$_{1}$-L$_{2}$                       & 71.1       & 5.0 (2.5--7.6)                            & 2.1                           & \textbf{0.038}                              \\
L$_{2}$-L$_{3}$                       & 45.3               & 3.0 (0.5--5.5)                     & 1.3                           & 0.182                                       \\
L$_{3}$-L$_{4}$                        & 145.7          & 11.6 (6.1--17.0)                         & 4.3                           & \textbf{\textless0.001}                     \\
L$_{4}$-L$_{5}$                     & 151.8              & 16.0 (5.0--26.9)                     & 4.5                           & \textbf{\textless0.001}                     \\
L$_{5}$-S$_{1}$                      & 15.4                & 1.1 (-7.2--9.4)                    & 0.5                           & 0.674                                       \\
&  &  &  &  \\
$ADC_{post}$                     &                   &                      &                               &                                             \\
L$_{1}$-L$_{2}$                      & 126.3             & 9.3 (4.0--14.6)                      & 4.2                           & \textbf{\textless0.001}                     \\
L$_{2}$-L$_{3}$                       & 76.1               & 6.0 (1.4--10.6)                     & 2.5                           & \textbf{0.013}                              \\
L$_{3}$-L$_{4}$                      & 131.0                & 10.1 (5.2--15.1)                  & 4.3                           & \textbf{\textless0.001}                     \\
L$_{4}$-L$_{5}$                       & 159.8                  & 20.1 (7.5--32.6)                & 5.3                           & \textbf{\textless0.001}                     \\
L$_{5}$-S$_{1}$                       & 59.1                      & 8.6 (-0.9--18.1)              & 1.9                           & 0.053                                       \\ \bottomrule
\end{tabular}
\end{table}

\subsection*{Relationships between clinical and $ADC$ results}
\begin{figure}[!h]
\centering
\includegraphics[scale=0.5]{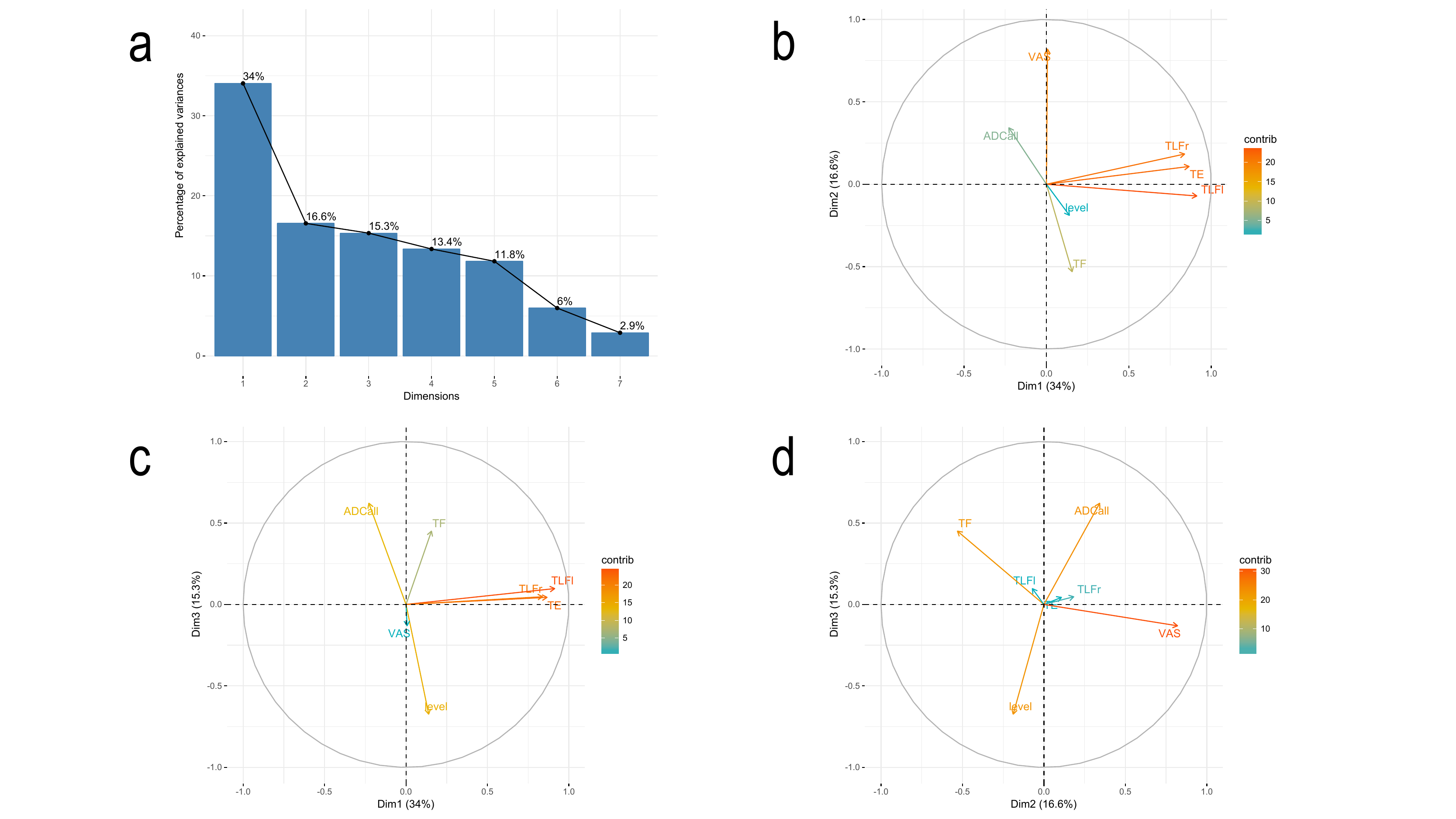}
\caption{(a) Scree plot of percentage of explained variances after PCA. (b) PCA results: correlation circle for dimensions 1 and 2. (c) PCA results: correlation circle for dimensions 1 and 3. (d) PCA results: correlation circle for dimensions 2 and 3.}
\label{fig:pca}
\end{figure}

PCA results are presented in Figure \ref{fig:pca}. Both the Kaiser \cite{Kaiser:1960} rule of eigenvalues greater than 1 (component 1=2.38, component 2=1.16, and component 3=1.07) and the scree plot \cite{Cattell:1966} of the percentage of explained variances by each of the components as a percentage of the total variance (see Figure \ref{fig:pca}a) indicated that three-factor solution fit the data the best, explaining a cumulative percentage of variance of 65.9\%.

PCA results are summarized in 3 correlation circles, with variable contribution to the principal axes (`contrib') coded in colors (Figures \ref{fig:pca}b to \ref{fig:pca}d). The main contribution of variables to dimension 1 were $\Delta TLF_{l}$, $\Delta TE$, and $\Delta TLF_{r}$. Dimension 2 was mainly explained by $\Delta VAS$ and $\Delta TF$, and dimension 3 by anatomical level, $\Delta ADC_{all}$, and $\Delta TF$. $\Delta VAS$ was negatively correlated with $\Delta TF$ (Figs. \ref{fig:pca}b and \ref{fig:pca}d) and $\Delta ADC_{all}$ with anatomical level (Figs. \ref{fig:pca}c and \ref{fig:pca}d). 

\section*{Discussion}
The rationale for studying an acute LBP population was based on previous research findings that subjects with longer than 2-month symptoms durations did not respond as well to a manual therapy mobilization \cite{Beattie:2010}. Second, even if MRI is a technique capable of providing information both on the morphology of the IVD and on its molecular composition, it is desirable to direct research effort toward characterizing changes that are linked directly to clinical symptoms \cite{Urban:2007}.

Our results support previous findings of a simultaneous pain reduction and increase of $ADC$ in the NP of chronic LBP subjects after PA lumbar mobilization \cite{Beattie:2010} but provide new data concerning the acute phase of disease, and trunk mobility in an older population with higher pain intensity levels. Beattie \textit{et al.} \cite{Beattie:2010} were the first to explore the short-term effect of oscillating PA pressures to the lumbar spinous processes followed by prone press-ups exercises in chronic LBP subjects on pain intensity and water diffusion within NP of IVD. They observed two subgroups: ``within-session responders'' and ``not-within-session responders'', based on a reduction of pain of at least 2/10 within-session or not. No attempt was made to separate our sample into ``within-session responders'' and ``not-within-session responders'' since its small size and that only 4 subjects show a pain reduction of less than 2/10, sometimes combined with a large increase in $ADC$ values.

Mean age of our population was 46 years with a pain intensity at baseline of 5.4/10 on VAS. The mean age of the population studied by Beattie \textit{et al.} \cite{Beattie:2010} was 26 years with an average pain intensity on a typical day of 3.7/10 on the 11-point numeric rating scale. The difference in pain intensity between the two studies could not be explained by gender differences, since 9/12 (75\%) subjects in the ``within-session responders'' group of Beattie's study were female and 11/16 (69\%) in ours. On the other hand, a difference in body mass index (BMI) could explain it, since higher values are associated with higher pain intensity levels in patients with LBP \cite{Ojoawo:2011, Hussain:2017}. A mean lower value of 21.0 $kg~m^{-2}$ was observed in ``within-session responders'' of Beattie's study compared to 26.6 in ours. 

The 62\% mean reduction in pain following PA mobilization is higher than that reported in previous investigations, with a mean decrease ranging between 33 and 41\%, when mobilization was applied: on the most painful lumbar level, at a random lumbar level, or even at painful lumbar level and all other lumbar levels \cite{Chiradejnant:2002, Goodsell:2000, Powers:2008}. A potential explanation of this difference may the lower homogeneity of the patient's groups of previous investigations that include LBP subjects with too long pain symptoms duration: up to 3 months \cite{Powers:2008}, more than 6 months \cite{Chiradejnant:2002}, and even up to 60 months \cite{Goodsell:2000}.

Normal IVD is considered as a poorly innervated organ, since its innervation is restricted to the outer layers and consists of small nerve fibers and some large fibers forming mechanoreceptors. Nerve fibers accompany the blood vessels or arrive via independent ways: branches of sinuvertebral nerve, nerve branches from the ventral rami of spinal nerves, or gray rami communicantes \cite{Bogduk:1983}. IVD could also receive nerve branches from the anterior and posterior longitudinal ligaments \cite{Bogduk:1983}. In contrast, in degenerative IVD, Coppes \textit{et al.} \cite{Coppes:1997} demonstrated a more important and profound innervation compared to normal discs. Furthermore, nociceptive properties of at least some of these nerves are strongly suggested by their immunoreactivity for substance P. These observations are used to defend the hypothesis of the existence of discogenic pain, in degenerative IVDs. By definition, discogenic pain is a pain due to a mechanical or chemical irritation of nerves innervating the IVD. Based on our results and those of Beattie and colleagues \cite{Beattie:2008, Beattie:2009, Beattie:2010, Beattie:2014}, we believe that the simultaneous reduction in pain observed in patients and increase of the water diffusion within IVD is not an epiphenomenon linked to mobilization, and that, on the contrary, these two physiological events would be intimately related, directly or indirectly. It is not inconsistent to speculate that an increased water diffusion would lead to a re-expansion of the IVD and therefore reduce the mechanical stresses on the large mechanoreceptors nerve fibers. Furthermore, increasing the speed of the water and blood flow in the IVD could decrease local inflammatory process and thus the pain.

On one side, it is accepted that onset of the disc degeneration process start to occur in the third decade of life, with dehydration of the NP and changes in the molecular structures of its components \cite{Nguyen:2008}. On the other side, a link exists between water diffusion in NP, estimated by $ADC$, and visual degeneration of lumbar IVD, using Pfirrmann's grading system \cite{Niinimaki:2009}. Surprisingly, a reduction in $ADC$ values of 4\% was observed between normal and moderately degenerated discs but severely degenerated discs showed 5\% larger $ADC$ values than normal discs, presumably due to free water in cracks and fissures in the degenerated NP of those discs \cite{Niinimaki:2009}. After a spinal thrust, LBP subjects with fewer lumbar degenerated discs showed better increased in $ADC$ values than those with many \cite{Beattie:2014}. Here, the majority of IVD graded as moderately degenerated for more cranial anatomical levels and as severely degenerated for more caudal levels, and $ADC$ changes were higher at more cranial levels compared to caudal, with non significant changes at L$_{5}$-S$_{1}$.

To our knowledge, changes in trunk mobility have never been studied concurrently with changes in pain and water diffusion within NP. Even if the assessment of trunk mobility is a strong point of our protocol, a potential bias is that the investigator that assess trunk mobility was not blinded to if PA had been performed or not. Using a principal component analysis (PCA), several novel and important observations were made about the relationships between changes in pain, trunk mobility and water diffusion. First, a negative correlation between changes in pain and changes in trunk flexion was observed, but not with changes in extension and lateral flexions. Second, a negative correlation between changes in water diffusion and lumbar anatomic levels was observed. In line with previous findings, the mobility of trunk in extension \cite{McCollam:1993, Shum:2013, Shah:2016} and in flexion \cite{Shum:2013} improved significantly after PA mobilization. However, some studies failed to report significant increase in trunk extension \cite{Chiradejnant:2002, Goodsell:2000} and flexion \cite{McCollam:1993}. We show a significant increase of 29.9$\pm$23\% for trunk flexion, 8.1$\pm$8\% for trunk extension, 9.9$\pm$8\% for left lateral trunk flexion, and 8.9$\pm$8\% for right lateral trunk flexion. The significant mean change of 9 $cm$ we observed for major fingertip-to-floor distance during trunk flexion after PA mobilization in our acute population, was greater than the significant mean change of 2.7 $cm$ reported by Goodsell \textit{et al.} \cite{Goodsell:2000} in chronic subjects. In contrast to the non-significant mean change of 0.3 and 0.12 $cm$ for right and left lateral trunk flexions reported by Samir \textit{et al.} \cite{Samir:2016} in chronic subjects after PA mobilization, we observed a significant mean change of 5 and 4 $cm$. Our results suggest that trunk mobility improvements after PA mobilizations could be more important in acute subjects than chronic. However, fingertip-to-floor method measures total forward, backward, and lateral bending movements, including movement of the spine, hips, and pelvis. This method does not allow to specify at which level mobility changes occur.

Although the use of DW MRI in humans has mainly been applied to the central nervous system and in particular the brain, more recently, this method has become increasingly successful in the musculoskeletal system and has led to a broadening of knowledge both in diagnosis and intervention, using the $ADC$. $ADC$ values were determined in 80 lumbar IVDs, from L$_{1}$-L$_{2}$ to L$_{5}$-S$_{1}$ levels. An increase in $ADC_{all}$ of 7.2\% was observed for L$_{1}$-L$_{2}$; 5.9\% for L$_{2}$-L$_{3}$; 13.2\% for L$_{3}$-L$_{4}$; 16.0\% for L$_{4}$-L$_{5}$ and 4.1\% for L$_{5}$-S$_{1}$. Beattie \textit{et al.} \cite{Beattie:2010} observed a mean $ADC$ increase of 4.2\% within L$_{5}$-S$_{1}$ IVD in `immediate responder' group (n=10) after PA mobilization. At all anatomical levels, change in $ADC_{all}$ values were greater than $SEM_{\%}$ of 2.1 observed on one subject after 10 minutes of prone lying, which is compatible to the SEM values reported by Beattie \textit{et al.} \cite{Beattie:2009} on 24 subjects after 10 minutes of prone lying and ranging from -3.5 to 3.4\%. Therefore, $ADC_{all}$ changes observed after PA mobilization must be considered as real changes linked to mobilization and not to measurement errors. Even if has been long established that the IVD is one of the largest avascular anatomical structure in the body \cite{Urban:2007}, it nevertheless remains a living structure that requires convection and diffusion mechanisms to ensure nutrition. Diffusion is defined as the movement of matter driven by a concentration gradient and convection is described as the bulk movement of fluids \cite{Jackson:2009}. It is generally believed that diffusion is the main transport mechanism for small solutes with convection playing a more important role in the transport of larger solutes \cite{Jackson:2009}. DW images  provide a characterization of water transport under the combined influence of diffusion and convection. An increase of diffusion/convection in the NP is thought to be beneficial, while decreased diffusion/convection has been linked with degeneration. Diffusion of water within the IVD is influenced by pressure gradients and chemical forces acting on it, as well as structural barriers such as a nuclear ``cleft''. Pressure gradients within IVD could be influenced by externally applied forces, such as those generated by manual therapy techniques \cite{Beattie:2014, Adams:2006, Ferrara:2005}. We hypothesize that diffusion of water could be related to opening-closure mechanism of IVD. This mechanism has been observed \textit{in vivo} by Kulig \textit{et al.} \cite{Kulig:2004}, when applying a PA pressure at the lumbar spine. A pressure applied at a given vertebral level results in an extension movement (opening) at this level and on the upper level, and on the contrary a movement of flexion (closure) on the lower level.

Correlations were previously described between anatomical levels and $ADC$ values but findings were inconsistent. Some studies show that $ADC$ values increase significantly with more caudal IVDs \cite{Kerttula:2000, Kealey:2005}, decrease significantly with more caudal IVDs \cite{Niu:2011}, or even are not significantly correlated with IVD levels \cite{Niinimaki:2009}. In a more recent study \cite{Wu:2013}, the influence of age on these relationships was observed, with $ADC$ mean values for young subjects ($<$45 years) increasing from L$_{1}$-L$_{2}$ to L$_{2}$-L$_{3}$/ L$_{3}$-L$_{4}$ levels and decreasing to more caudal levels, and decreasing continuously for elderly subjects ($>$45 years). Furthermore, static traction was associated with an increase in diffusion of water within the L$_{5}$-S$_{1}$ IVDs of middle-age individuals, but not in young adults, suggesting age-related differences in the diffusion response \cite{Mitchell:2017}. Here, PCA results show that $ADC_{all}$ values tend to decrease with more caudal IVDs. 

Today, there is a paucity of research that describes the physiologic events associated with analgesia following intervention for LBP \cite{Beattie:2014}. Since $ADC$ is a measure of the magnitude of random (Brownian) diffusion motion of water molecules, it provides information about the physiologic state of the NP. Previous studies estimate $ADC$ of NP with only one ROI. Here, $ADC_{all}$ was estimated from the mean of anterior, middle, and posterior portions of the NP, which were themselves estimated based on the mean of 3 ROIs (sagittal medial, and left and right parasagittal planes). We believe that our method is more representative of a physiological/ physiopathological process of the entire NP than measures based on a single ROI analysed in the mid-sagittal scan, since pathologically relevant disc measurements may be observed in parasagittal or other planes \cite{Violas:2005}.

Greatest changes in $ADC_{all}$ were observed at L$_{3}$-L$_{4}$ and L$_{4}$-L$_{5}$ levels, and are mainly explained by changes in $ADC_{ant}$ and $ADC_{post}$. Note that PA mobilizations were applied between L$_{3}$ and L$_{5}$ in 15 subjects on 16. Since $ADC_{ant}$ and $ADC_{post}$ were greater than $ADC_{mid}$ changes, and taken together with pain decrease, our results suggest that increased peripheral random motion of water molecules in nucleus pulposus is implicated in the modulation of the IVD nociceptive response. This observation is all the more important since nerve fibres have been identified in the NP of degenerated IVDs \cite{Binch:2015}, which may still be more likely to be able to generate an efficient reduction of pain than healthy IVDs that are usually thought to be innervated only in the annular part. Therefore, it would be interesting to study the influence of these mobilizations, both in nucleus pulposus and annulus fibrosus, according to the 3 orthogonal directions of space (x,y,z) rather than using an average value of $ADC$. Pure water, for the purposes of diffusion is said to be isotropic; this means that the molecules are equally likely to diffuse in any direction. In a biological tissue like the NP, there may be a preferential diffusion direction, along collagen fibers, and diffusion is said anisotropic. Our methodology does not allow to study the anisotropic character of water diffusion within NP. This latter has already been observed previously within lumbar IVDs on healthy young adults \cite{Kerttula:2000}, with $ADC_{z}$ (diffusion perpendicular to the end-plate) values higher than $ADC_{x}$ and $ADC_{y}$ (diffusion in the disc plane). Very recently, a promising T2-weighted MRI method based on the location of the signal intensity weighted centroid, i.e. the arithmetic mean of the signal intensity of all pixels in a ROI, was developed as a biomarker for investigating fluid displacement within the disc \cite{Abdollah:2017}. It would be interesting to apply this method to our images. 

From L$_{1}$-L$_{2}$ to L$_{5}$-S$_{1}$ IVD levels, the mean NP length in the sagittal plane is comprised between 19.3$\pm$2.9 and 21.6$\pm$3.1~$mm$, and height between 5.5$\pm$1.1 and 8.6$\pm$1.3~$mm$ \cite{Zhong:2014}. The method used here was based on the use of ROIs always having a circular surface area of 0.2 $cm^{2}$, either a diameter of 5~$mm$, resulting in a total surface area of 15~$mm$ long (anterior, middle, and posterior ROIs) by 5~$mm$ high in the sagittal and the two parasagittal planes. Elliptical surfaces of varying dimensions, ranging from 40 and 80~$mm^{2}$, have been used by others \cite{Kealey:2005, Niu:2011, Belykh:2017}, forcing the observers to place the large axis in the ventro-dorsal direction and the small axis in the cranio-caudal direction. The risk of using surfaces up to 80~$mm^{2}$ is to include the most internal part of the AF in the calculation of the $ADC$. 

This study was limited by the absence of a T1-weighted MRI sequence in order to estimate vertebral endplate signal changes and classify it according to their levels of degeneration  \cite{Modic:1988}. Indeed, there is strong evidence that vertebral endplate structural changes are associated with non-specific LBP but it may be present in individuals without LBP \cite{Jensen:2008}. Since the main and most important pathway for diffusion into the NP occurs from capillaries in the vertebral body via diffusion through the cartilaginous endplate \cite{van_der_Werf:2007}, another limitation is the lack of evaluation of vertebral endplate morphology. As described by Lakshmanan \textit{et al.} \cite{Lakshmanan:2012}, concavity of the lumbar endplates is symmetrical in the frontal plane but shape shows considerable variability in the sagittal plane (flat, oblong or ex-centric), with inferior endplate shape becoming more ex-centric, i.e. location of the concavity apex in the posterior half of endplate (54--60\% endplate diameter), from L$_{3}$ to L$_{5}$ levels. At these levels, significant $ADC$ changes were observed within NP, corresponding approximately to the center or apex of the endplate, suggesting that the mechanical stimuli induced by PA mobilization may have a direct influence on vertebral endplates. By the way, permeability across the cartilage end plate is greater in the central portion, adjacent to the NP, than at the periphery, near the AF \cite{Roberts:1989}. Finally, no attempt was made to assess subject's functional disability; the Oswestry Disability Index \cite{Fairbank:2000}, considered as the gold standard for measuring degree of disability and estimating quality of life in a subject with LBP, could have been realized to complete the clinical picture of our sample.

\bibliography{adc}

\noindent

\section*{Acknowledgements}
The authors would like to thank Siemens and Grand H\^{o}pital de Charleroi for the financial support provided for the production of magnetic resonance imaging scans.

\section*{Author contributions statement}
T.P., R.F. and D.F. conceived and conducted the experiments, R.F. and D.F. analysed the results. All authors reviewed the manuscript. 

\section*{Additional information}


The authors declare no competing financial interests.

\end{document}